\documentclass[twocolumn]{aastex631}
\usepackage{CJKutf8} 
\usepackage{amsmath}
\usepackage{comment}
\usepackage{physics}
\usepackage{verbatim}
\usepackage{hyperref}
\usepackage{tablefootnote}
\usepackage[T1]{fontenc}
\usepackage[utf8]{inputenc}

\usepackage{soul}

\submitjournal{AJ}

\begin{document}

\title{TOI-5375 B: A Very Low Mass Star at the Hydrogen-Burning Limit Orbiting an Early M-type Star\footnote{Based on observations obtained with the Hobby-Eberly Telescope (HET), which is a joint project of the University of Texas at Austin, the Pennsylvania State University, Ludwig-Maximillians-Universitaet Muenchen, and Georg-August Universitaet Goettingen. The HET is named in honor of its principal benefactors, William P. Hobby and Robert E. Eberly.}\footnote{The WIYN Observatory is a joint facility of the NSF's National Optical-Infrared Astronomy Research Laboratory, Indiana University, the University of Wisconsin-Madison, Pennsylvania State University, the University of Missouri, the University of California-Irvine, and Purdue University.} }
  
\correspondingauthor{Mika Lambert}
\email{mlambert43@arizona.edu}

\author[0000-0002-2527-8899]{Mika Lambert}
\affiliation{Steward Observatory and Department of Astronomy, University of Arizona, 933 N. Cherry Ave., Tucson, AZ 85721, USA}

\author[0000-0003-4384-7220]{Chad F. Bender}
\affiliation{Steward Observatory and Department of Astronomy, University of Arizona, 933 N. Cherry Ave., Tucson, AZ 85721, USA}

\author[0000-0001-8401-4300]{Shubham Kanodia}
\affiliation{Earth and Planets Laboratory, Carnegie Institution for Science, 5241 Broad Branch Road, NW, Washington, DC 20015, USA}

\author[0000-0003-4835-0619]{Caleb I. Cañas}
\affiliation{NASA Goddard Space Flight Center, 8800 Greenbelt Road, Greenbelt, MD 20771, USA}
\affiliation{Department of Astronomy \& Astrophysics, 525 Davey Laboratory, The Pennsylvania State University, University Park, PA, 16802, USA}
\affiliation{Center for Exoplanets and Habitable Worlds, 525 Davey Laboratory, The Pennsylvania State University, University Park, PA, 16802,
USA}

\author[0000-0002-0048-2586]{Andrew Monson}
\affiliation{Steward Observatory and Department of Astronomy, University of Arizona, 933 N. Cherry Ave., Tucson, AZ 85721, USA}

\author[0000-0001-7409-5688]{Guðmundur Stefánsson}
\affil{NASA Sagan Fellow}  
\affil{Department of Astrophysical Sciences, Princeton University, 4 Ivy Lane, Princeton, NJ 08540, USA}

\author[0000-0001-9662-3496]{William D. Cochran}
\affiliation{McDonald Observatory and Department of Astronomy, The University of Texas at Austin, USA}
\affiliation{Center for Planetary Systems Habitability, The University of Texas at Austin, USA}
\author[0000-0002-0885-7215]{Mark E. Everett}
\affiliation{NSF’s National Optical-Infrared Astronomy Research Laboratory}
\author[0000-0002-5463-9980]{Arvind F. Gupta}
\affil{Department of Astronomy \& Astrophysics, 525 Davey Laboratory, The Pennsylvania State University, University Park, PA, 16802, USA}
\affil{Center for Exoplanets and Habitable Worlds, 525 Davey Laboratory, The Pennsylvania State University, University Park, PA, 16802, USA}
\author{Fred Hearty}
\author[0000-0002-4475-4176]{Henry A. Kobulnicky}
\affiliation{Department of Physics \& Astronomy, University of Wyoming, Laramie, WY 82070, USA}
\author[0000-0002-2990-7613]{Jessica E. Libby-Roberts}
\affiliation{Department of Astronomy \& Astrophysics, 525 Davey Laboratory, The Pennsylvania State University, University Park, PA, 16802, USA}
\affiliation{Center for Exoplanets and Habitable Worlds, 525 Davey Laboratory, The Pennsylvania State University, University Park, PA, 16802,
USA}
\author[0000-0002-9082-6337]{Andrea S.J.\ Lin}
\affil{Department of Astronomy \& Astrophysics, 525 Davey Laboratory, The Pennsylvania State University, University Park, PA, 16802, USA}
\affil{Center for Exoplanets and Habitable Worlds, 525 Davey Laboratory, The Pennsylvania State University, University Park, PA, 16802, USA}

\affiliation{Center for Exoplanets and Habitable Worlds, 525 Davey Laboratory, The Pennsylvania State University, University Park, PA, 16802,
USA}
\author[0000-0001-9596-7983]{Suvrath Mahadevan}
\affiliation{Department of Astronomy \& Astrophysics, 525 Davey Laboratory, The Pennsylvania State University, University Park, PA, 16802, USA}
\affiliation{Center for Exoplanets and Habitable Worlds, 525 Davey Laboratory, The Pennsylvania State University, University Park, PA, 16802,
USA}
\affiliation{ETH Zurich, Institute for Particle Physics \& Astrophysics, Switzerland}
\author[0000-0001-8720-5612]{Joe P. Ninan}
\affiliation{Department of Astronomy and Astrophysics, Tata Institute of Fundamental Research, Homi Bhabha Road, Colaba, Mumbai 400005,
India}
\author[0000-0001-9307-8170]{Brock A. Parker}
\affiliation{Department of Physics \& Astronomy, University of Wyoming, Laramie, WY 82070, USA}
\author[0000-0003-0149-9678]{Paul Robertson}
\affiliation{Department of Physics \& Astronomy, University of California Irvine, Irvine, CA 92697, USA}
\author[0000-0002-4046-987X]{Christian Schwab}
\affiliation{School of Mathematical and Physical Sciences, Macquarie University, Balaclava Road, North Ryde, NSW 2109, Australia}

\author[0000-0002-4788-8858]{Ryan C. Terrien}
\affiliation{Carleton College, One North College St., Northfield, MN 55057, USA}

\begin{abstract}
The TESS mission detected a companion orbiting TIC 71268730, categorized it as a planet candidate, and designated the system TOI-5375. Our follow-up analysis using radial velocity data from the Habitable-zone Planet Finder (HPF), photometric data from Red Buttes Observatory (RBO), and speckle imaging with NN-EXPLORE Exoplanet Stellar Speckle Imager (NESSI) determined that the companion is a very low mass star (VLMS) near the hydrogen-burning mass limit with a mass of 0.080$\pm{0.002} M_{\Sun}$ ($83.81\pm{2.10} M_{J}$), a radius of 0.1114$^{+0.0048}_{-0.0050} R_{\Sun}$ (1.0841$^{0.0467}_{0.0487} R_{J}$), and brightness temperature of $2600\pm{70}$ K. This object orbits with a period of 1.721553$\pm{0.000001}$ days around an early M dwarf star ($0.62\pm{0.016}M_{\Sun}$). TESS photometry shows regular variations in the host star's TESS light curve, which we interpreted as activity-induced variation of $\sim$2\%, and used this variability to measure the host star's stellar rotation period of 1.9716$^{+0.0080}_{-0.0083}$ days. The TOI-5375 system provides tight constraints on stellar models of low-mass stars at the hydrogen-burning limit and adds to the population in this important region. 
\end{abstract}


\section{Introduction}
\label{sec:intro}
The Transiting Exoplanet Survey Satellite \citep[TESS;][]{Ricker_2015} is a NASA mission to monitor nearly the entire sky for brief decreases in brightness caused by transiting planetary objects. However, a significant number of these transits are astrophysical false positives, caused by stellar binary systems. Since the launch of TESS in 2018, there have been 234 confirmed planets and 1573 false positives detected \citep{2021_TESS_cat}.
Ground-based follow-up is essential to fully characterize these objects.

Eclipsing binary systems are important astrophysical benchmarks because they allow us to dynamically constrain the physical characteristics of the system including mass and radius \citep[e.g.,][]{Torres_2009, Kesseli_2019, Serenelli_2021} mostly independent of theoretical models.
Thus, these stellar systems provide measurements that feedback into the calibration and evolution of stellar evolution models. Cataloging false positives in TESS data may also benefit the TESS data processing pipeline to identify parameters that are correlated with erroneously classifying binary systems as exoplanets.

The TESS input catalog identified TIC 71268730 (TOI-5375, 2MASS J07350822+7124020, Gaia DR3 1110586978339817728) as an early M dwarf with an effective temperature of $3865\pm{157}$ K. The TESS data processing pipeline designated the companion of TOI-5375 as a candidate planet with a period of 1.72 days and a depth of $36.88\pm{0.58}$ mmag. \cite{Gan_2022} classify TOI-5375 as a verified planet candidate after vetting by their photometric analysis pipeline.
In this paper, we present our analysis of the TOI-5375 system augmenting the TESS photometry with ground-based observations to determine the companion, TOI-5375 B, is a very low mass star (VLMS) at the hydrogen-burning mass limit. In \S \ref{sec:method}, we describe the observational data collected; in \S \ref{sec:Stellar Parameters}, we discuss the stellar parameters; in \S \ref{sec:analysis}, we discuss the resulting posteriors of our joint fit; in \S \ref{sec:Discussion} we present an analysis of our results in the context of evolutionary models, age, temperature, and environment.

\section{Observations}
\label{sec:method}
\subsection{TESS Photometry}
TOI-5375 was observed by TESS in Sector 20 from 2019 December 24 - 2020 January 21, and in Sector 26 from 2020 June 8 - 2020 July 4, at a thirty-minute (1800 seconds) cadence. It was also observed in Sector 40 from 2021 June 24 - 2021 July 23, 
at 120-second cadence. 
Similar to the TOI-1899 \citep{Canas_2020} and TOI-3629 \citep{Canas_2022} systems, we identified TOI-5375 B as a planetary candidate using a custom pipeline to search for transiting candidates in short and long-cadence TESS data orbiting M dwarfs that were amenable to RV observations with HPF \citep[see;][]{Canas_2022}. TOI-5375 B was also independently identified by the TESS science processing pipeline \citep{SPOC} with a period of about 1.72 days and a transit duration of 1.74 hours. For the short cadence data, Sector 40, we obtained the Pre-search Data Conditioning SAP flux (PSDCSAP) data from the Mikulski Archive for Space Telescopes (MAST).

\begin{figure*}
    \centering
    \includegraphics[width=\textwidth]{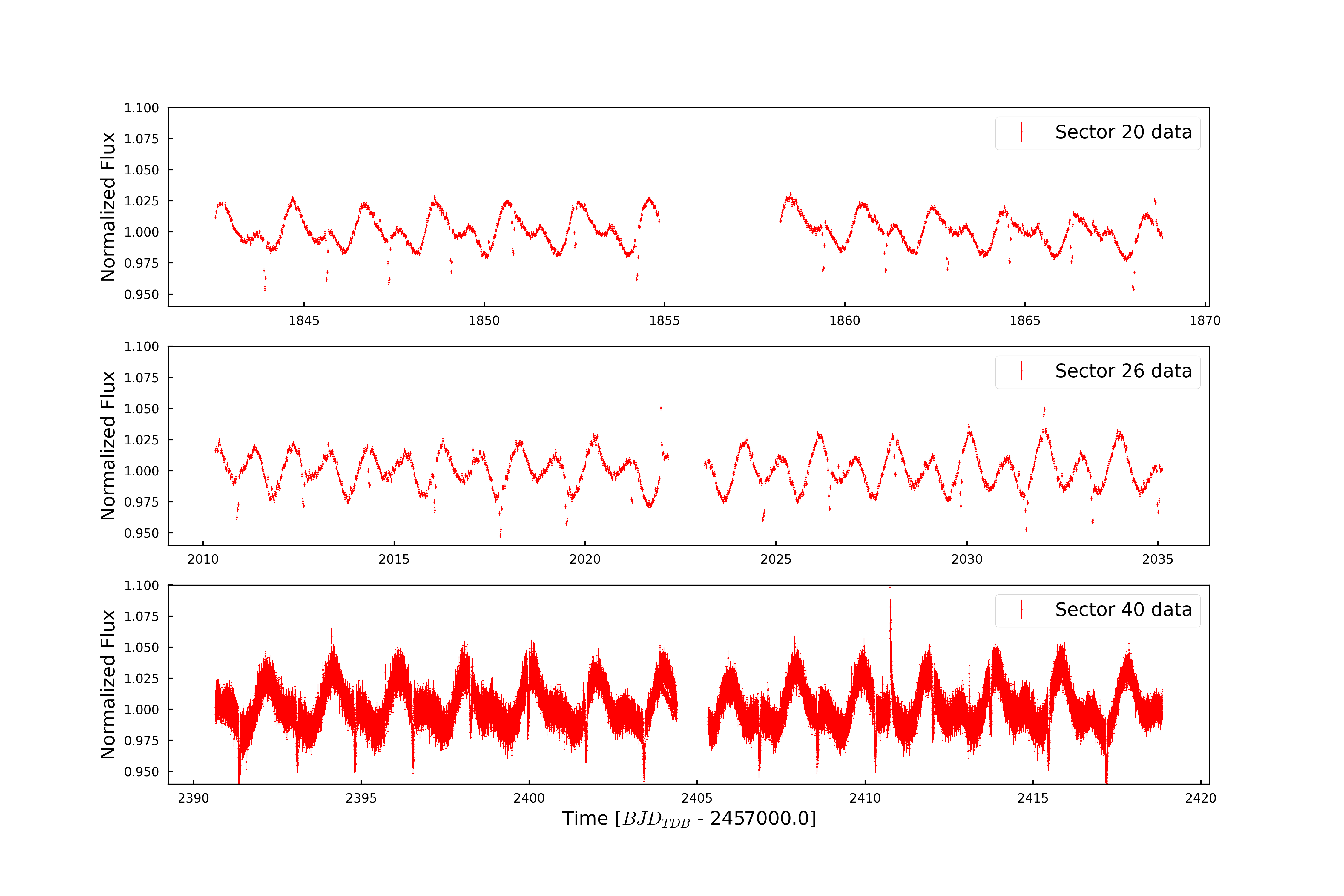}
    \caption{Long cadence (1800 seconds), unbinned transit observations of TOI-5375 in TESS Sector 20 (top), 26 (middle), and short cadence (120 seconds) unbinned photometry from TESS Sector 40 (bottom). Variability in the host star's light curve evolves throughout the different sectors which we infer to be due to varying stellar spots. The periodicity of the spot-induced variability is tied to the rotation of the primary star. We also see flares due to stellar activity, which are masked out in our analysis.}
    \label{fig:lc}
\end{figure*}

We used $\tt{eleanor}$ \citep{eleanor2019} to produce the light curves from the TESS full-frame images of Sectors 20 and 26. $\tt{eleanor}$ uses the TESScut\footnote{\url{ https://mast.stsci.edu/tesscut/}} service to obtain a cut-out of 31$\times$31 pixels from the calibrated full-frame images centered on the target. In order to derive the light curve, we used the $\tt{CORR FLUX}$ values, in which $\tt{eleanor}$ uses linear regression with pixel position, measured background, and time to remove signals correlated with these parameters. We set the aperture mode to `normal' which tests different apertures and is based on the magnitude of the target star and the contamination ratio from TESS \citep{eleanor2019}. Figure \ref{fig:lc} shows the original light curves of TOI-5375 in TESS sector 20, 26, and 40. Several strong flares are clearly seen in the light curves. We identified and masked these events by hand before carrying out subsequent analysis.

\subsection{Ground Based Follow Up}
\subsubsection{RBO Photometry}

We observed TOI-5375 on the night of 2022, April 4 UT using the 0.6 m telescope at Red Buttes Observatory (RBO) in Wyoming \citep{ 2016_RBO_night_obs}. RBO is equipped with an Andor Apogee Alta F16 camera and used the 2×2 on-chip binning mode, which has a gain of 1.4 e$^-$/ADU, and a plate scale of 0.73"/pixel. All observations were obtained in the Bessell I filter \citep{Bessell_1990}. The target was defocused moderately and observed using an exposure time of 240 seconds. Observations ranged from an airmass of 1.18 to 2.15. 
We processed the RBO light curves using AstroImageJ \citep{Collins_2017}. The final reductions used a photometric aperture radius of 12 pixels (8.76"), an inner sky radius of 18 pixels (13.14"), and an outer sky radius of 25 pixels (18.25").

\subsubsection{HPF Radial Velocities}
From November 2020 to January 2022, we used the Habitable-zone Planet Finder \citep[HPF;][]{HPF2014} to obtain 12 exposures of TOI-5375. HPF is a high-resolution, near-infrared (8080 $-$ 12780 {\AA}) Doppler spectrograph at the 10-meter Hobby-Eberly Telescope (HET) located in Texas \citep{HET_1998, HET}. We used the tool $\tt{HxRGproc}$ to convert the raw HPF data into flux images and correct nonlinearity and cosmic rays, remove bias noise, and calculate the slope/flux and variance image \citep{Ninan2018}.

We analyzed the HPF spectra to measure the radial velocities (RVs) using the method in \cite{HPF_Stefansson}, which uses a modified version of the $\tt{SpEctrum}$ $\tt{Radial}$ $\tt{Velocity}$ $\tt{AnaLyser}$ pipeline ($\tt{SERVAL}$) \citep{Serval2018} that has been optimized for HPF data.
HPF-adapted $\tt{SERVAL}$ first creates a master template from the target star observations and then moves it in velocity space to determine the Doppler shift for each observation. $\tt{SERVAL}$ then compares the observation with the template and minimizes the $\chi^2$ statistic. The telluric regions are identified by a synthetic telluric-line mask created by $\tt{telfit}$ \citep{telfit}, a Python wrapper to the Line-by-Line Radiative Transfer Model package \citep{clough2005}. After masking out the telluric and sky-emission lines, the master template is created using all of the HPF observations for this target. We used $\tt{barycorrpy}$ \citep{Kanodia_2018} to account for the barycentric correction on each spectra. The RVs, 1$\sigma$ RV uncertainty, S/N, and exposure times are listed in Table \ref{tab:RV table}.

\begin{deluxetable}{lcrrr}
\tablecaption{HPF observations of TOI-5375.}
\label{tab:RV table}
\tablewidth{0pt}
\tablehead{
\colhead{BJD\tablenotemark{1}}&\colhead{RV (m/s)}&\colhead{$\sigma$ (m/s)}&\colhead{SNR\tablenotemark{2}}\\}
\startdata
2459159.972591&	1077.82&	95.49&  57\\
2459159.984070&	1458.72&	101.72& 55\\
2459159.995844&	2260.30&	93.86&  60\\
2459268.667101&	9694.40&	140.97&    42\\
2459268.675136&	9903.73&	161.72& 37\\
2459268.683141&	9943.74&	117.58&    50\\
2459517.999874&	-2325.99&	115.32& 50\\
2459520.006804&	9021.04&	109.39&    52\\
2459530.957075&	-17331.27&	82.48&    66\\
2459538.950183&	10048.02&	124.83&    48\\
2459547.928928&	-1769.31&	92.52& 60\\
2459596.774456&	-25396.43&	87.53&    63\\
\enddata
\tablenotetext{1}{BJD is the Barycentric Julian Date.}
\tablenotetext{1}{The SNR is the sn18 value which is the median SNR in order 18 at 1070 nm. The exposure times are 945 seconds except for 2459268.*, which are 630 seconds. which are 630 seconds.}
\end{deluxetable}

\subsection{NESSI Speckle Imaging}
\begin{figure}[ht]
    \centering
    \includegraphics[scale=0.43]{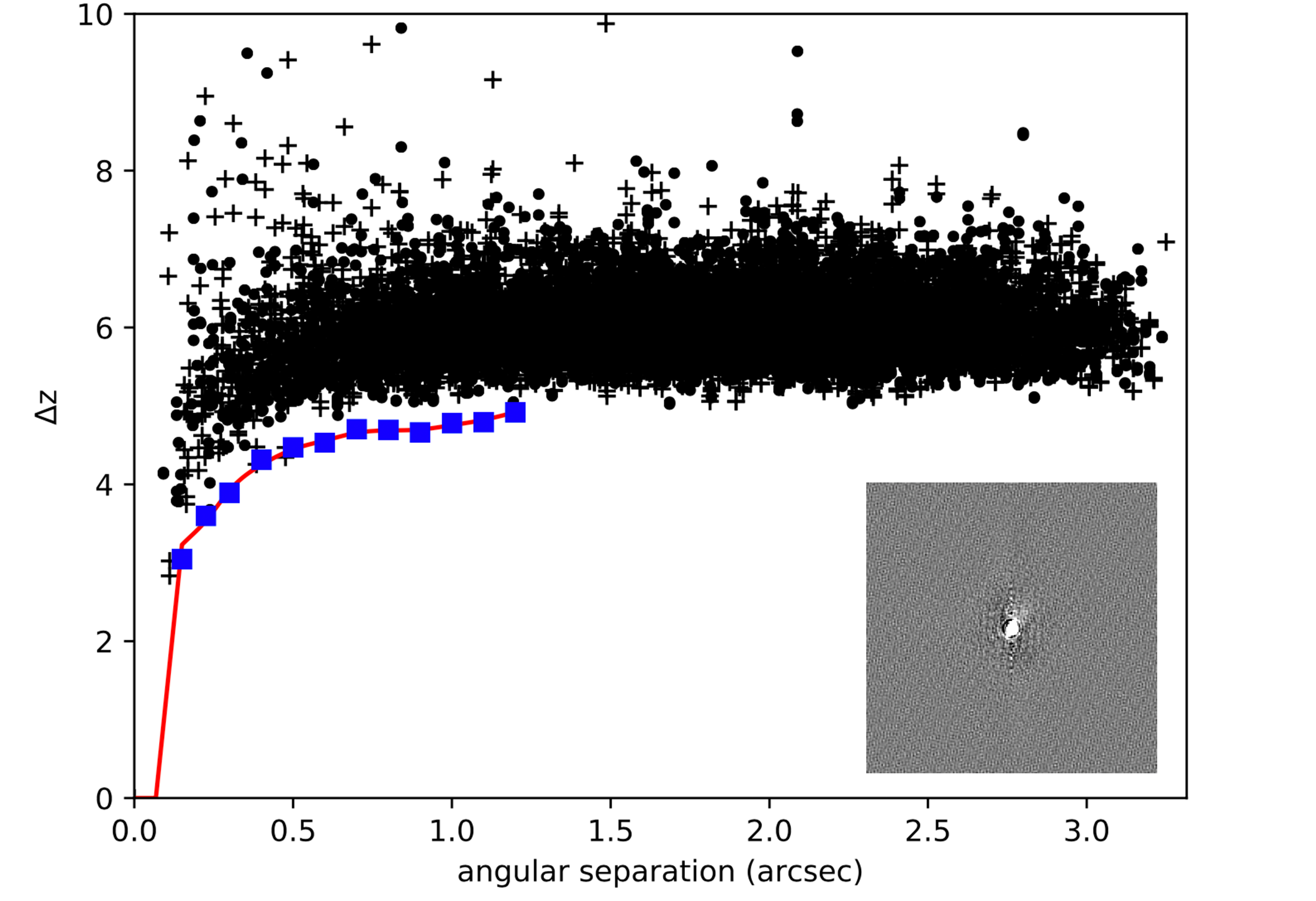}
    \caption{Speckle contrast curve of TOI-5375 from the Sloan z' filter using NESSI. The blue squares are 5-sigma contrast limits at each angular separation with a spline fit in red. The black plus signs are the extreme local maxima and the black dots are extreme local minima for the sensitivity limits. The data reveals no bright companions and no significant sources of dilution at separations from 0.2" to 1.2" from TOI-5375. The inset is the 4.7" × 4.7" NESSI speckle image centered on TOI-5375 in the z' filter.}
    \label{fig:speckle}
\end{figure}

To investigate the possibility of bright background sources contaminating our RBO photometry, we observed TOI-5375 with the NN-Explore Exoplanet Stellar Speckle Imager \citep[NESSI;][]{Scott2018} on the WIYN 3.5m telescope
at Kitt Peak National Observatory on the night of 2022 April 21. A 9-minute sequence of 40 ms diffraction-limited speckle images was taken in the Sloan z' filter with NESSI's red camera. A reconstructed speckle image was generated following the procedures described in \cite{Howell2011}. Figure \ref{fig:speckle} shows the contrast curve along with an inset of the NESSI speckle image in the z' filter. We conclude that there are no close by sources with magnitudes brighter than $\Delta$z' = 4.45 for separations $>$ 0.5 arcsec.

\begin{deluxetable*}{lcrrr}
\tablecaption{Priors used in the joint fit. }
\label{tab:priors}
\tablewidth{0pt}
\tablehead{
\colhead{Parameter}&\colhead{Description}&\colhead{Model$^\tablenotemark{1}$}}
\startdata
Orbital Parameters:\\
~~P & Orbital Period (days)& $\mathcal N(1.72154439, 0.1)$\\
~~$T_0$ & Transit Midpoint ($BJD_{TDB}$) & $\mathcal N(2459391.4, 0.1)$ \\
~~log($R_B$/$R_A$) & Scaled Radius $ $ & $\mathcal L(-1.66211817, 1)$  \\
~~log(K) & RV semi-amplitude (m/s) & $\mathcal U(0, 10)$ \\
~~b & Impact Parameter & $\mathcal U(0, 1)$ \\
Other Constraints:\\
~~$R_{A}$ &  Stellar radius ($R_{\Sun}$)& $\mathcal N(0.649, 0.024)$ \\
~~$M_{A}$ &  Mass of star ($M_{\Sun}$)&  $\mathcal N(0.62, 0.016)$ \\
~~$M_{B}$ & Mass of companion ($M_{\Earth}$)& $\mathcal U(0.1, 3*10^6)$\\
~~$q$ & Mass ratio & $\mathcal U(0, 1)$ \\
~~ $S$& Surface brightness ratio& $\mathcal U(0, 1)$\\
~~$T_{eff_A}$ & Effective temperature of the host star (Kelvin)& $\mathcal N(3897, 88)$ \\
Jitter and Instrumental Terms:\\
~~$\gamma$ & Gamma velocity ($m$ $s^{-1}$) & $\mathcal N(1859, 20000)$\\
~~$u_1$ & Limb-darkening parameter$^\tablenotemark{2}$ & $\mathcal U(0, 1)$  \\ 
~~$u_2$ & Limb-darkening parameter$^\tablenotemark{2}$ & $\mathcal U(0, 1)$  \\ 
~~$dv/dt$ & HPF RV trend ($mm$ $s^{-1}$ $year^{-1}$)& $\mathcal N(0, 5)$ \\
~~sigma$_{RV}$ & RV jitter($m$ $s^{-1}$) & $\mathcal N(10^{-3},10^3)$\\
~~$D_{TESS}$ & TESS dilution & $\mathcal U(0.1, 1.5)$ \\
~~Q & Quality factor for secondary oscillation & $\mathcal U(0.01, 500.0)$\\
~~dQ & Difference between quality factor for primary and secondary model&  $\mathcal U(0.01, 500.0)$\\
~~f &Fractional amplitude of secondary&$\mathcal U(0.01, 1.0)$ \\
~~$log(\sigma_{phot})$ & Log jitter & $\mathcal U(-6, 1)$\\
\enddata
\tablenotetext{1}{$\mathcal N$ is normal, $\mathcal U$ is uniform, $\mathcal L$ is log normal.}
\tablenotetext{2}{Each object in the binary system has an independent limb-darkening parameter associated with it. The limb-darkening parameters are used in the secondary eclipse function.}
\end{deluxetable*}

\begin{deluxetable*}{lcrrr}
\tablecaption{Summary of the primary star's stellar parameters. }
\label{tab:stellar_param}
\tablewidth{0pt}
\tablehead{
\colhead{Parameter$^\tablenotemark{1}$}&\colhead{Description}&\colhead{Model}}
\startdata
~~$R_{A}$ & Stellar radius ($R_\Sun$) & $0.649\pm{0.024}$ \\
~~$M_{A}$ &Stellar mass ($M_\Sun$) & $0.620\pm{0.016}$\\
~~$\rho_A$ & Density (cgs)& $3.83^{+0.28}_{-0.24}$\\
~~A$_v$& V-band extinction (mag)& $0.017^{+0.014}_{-0.012}$\\
~~$T_{eff}$ (K)& Effective temperature (K) & $3897 \pm{88}$\\
~~[Fe/H] & Metallicity (dex)& $0.29\pm{0.12}$\\
~~log(g) & Surface gravity (cgs) &$4.68\pm{0.046}$\\
~~$v\sin i_A$ & Rotational broadening ($km~ s^{-1}$)&$ 16.7\pm{0.9}$\\
~~d& distance (pc) & $121.14 \pm 0.20$\\
\enddata
\tablenotetext{1}{Stellar parameters are derived from $\tt{HPF-SpecMatch}$.}
\end{deluxetable*}

\section{Stellar Parameters}
\label{sec:Stellar Parameters}
$\tt{HPF-SpecMatch}$ \citep{HPF_Stefansson} uses the empirical template matching methodology discussed in \cite{Yee_2017} to derive stellar parameters of the host star from HPF spectra. We used this package to calculate the stellar parameters effective temperature ($T_{eff}$), surface gravity (log(g)), metallicity ([Fe/H]), and $v\sin i_A$. $\tt{HPF-SpecMatch}$ identifies the spectra that best match well-characterized stars from a library using $\chi^{2}$ minimization. 
Then, it creates a composite spectrum using a weighted, linear combination of the five best-matching library spectra and derives the stellar properties using these weights.
While searching for the best-matching library spectra, $\tt{HPF-SpecMatch}$ uses a linear limb darkening law to broaden the stellar templates. We determined TOI-5375 has a $T_{eff}$ of $3897\pm{88}$ K, a log(g) of $4.68\pm{0.046}$, a [Fe/H] of $0.29\pm{0.12}$, and $v\sin i_A$ of $16.7\pm0.9$ km/s.
The reported uncertainty is the standard deviation of the residuals from a leave-one-out cross-validation procedure applied to the entire spectral library in the chosen spectral order.

\begin{figure*}[h]
    \centering
    \includegraphics[width=\textwidth,height=5cm]{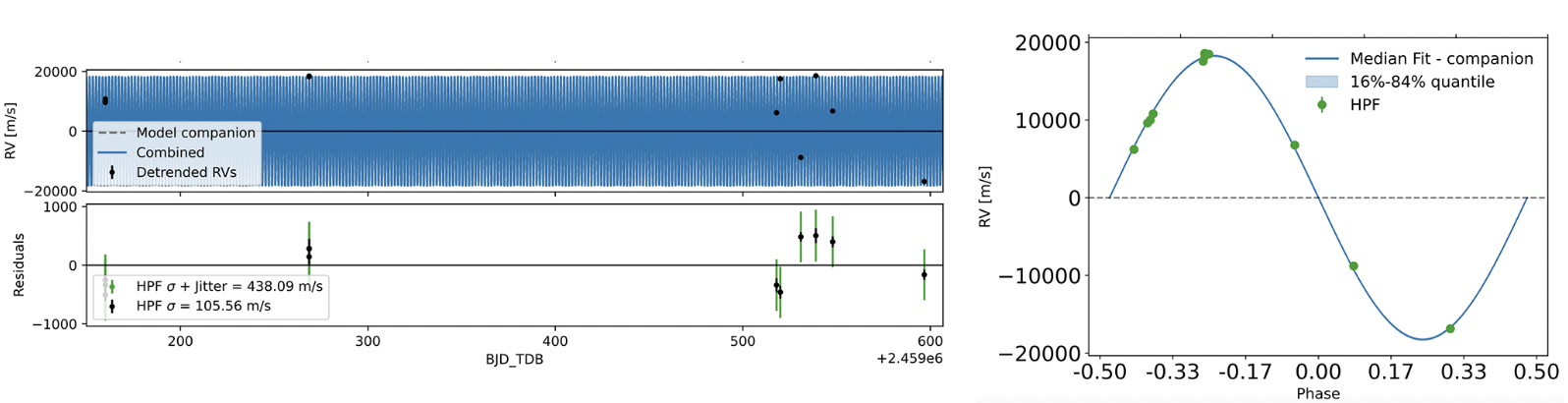}
    \caption{Top-Left: RV data points in black with the 
    best-fitting joint fit model overlaid in blue. Bottom-Left: Residuals from the best-fit model. Right: Phase folded RVs with the best-fitting joint fit model overlaid. }
    \label{fig:RV mcmc total}
\end{figure*}

We derived the model-dependent stellar parameters, mass and radius, using the spectral energy distribution (SED) which uses the $\tt{EXOFASTv2}$ analysis package \citep{Exofast}. $\tt{EXOFASTv2}$ calculates the bolometric corrections for the SED fit by linearly interpolating the precomputed bolometric corrections\footnote{\url{http://waps.cfa.harvard.edu/MIST/model_grids.html##bolometric}} of log(g), $T_{eff}$, [Fe/H], and AV from the MIST model grids \citep{2016_dotter, Choi_2016}. The SED fit uses Gaussian priors on: (i) 2MASS J, H, K magnitudes, Sloan g', r', i' magnitudes and Johnson B, V magnitudes from Henden et al. (2018), and Wide-field Infrared Survey Explorer magnitudes \citep{Wright_2010}; (ii) log(g), $T_{eff}$, and [Fe/H] derived from $\tt{HPF-SpecMatch}$, and (iii) the geometric distance calculated from \cite{Bailer_Jones_2021}. 

We applied an upper limit to the visual extinction based on estimates of Galactic dust \citep{Green_2019} calculated at the distance determined by \cite{Bailer_Jones_2021}. We converted the extinction from \cite{Green_2019} to a visual magnitude extinction using the Rv = 3.1 reddening law from \cite{1999_fitzpatrick}). Table \ref{tab:priors} contains the priors used in the joint fit described in Section \ref{sec:analysis}, and Table \ref{tab:stellar_param} contains the stellar parameters derived from our HPF SpecMatch analysis with their uncertainties. The model-dependent mass and radius are 0.649 $\pm{0.024}$ $M_\Sun$ and 0.620 $\pm{0.016}$ $R_\Sun$ respectively.

\section{Data Analysis}
\label{sec:analysis}

\subsection{Joint Fitting with Photometry and RV Data}
We used the $\tt{exoplanet}$ modeling code \citep{foremanMackey2021} to carry out a joint fit of TESS, RBO, and HPF data. $\tt{exoplanet}$ implements a Hamiltonian Monte Carlo (HMC) parameter estimation from PyMC3 \citep{pymc3} using the Gelman–Rubin statistic of $\hat{R}\leq1.1$ \citep{ford} to check for convergence.

$\tt{exoplanet}$ uses $\tt{starry}$ \citep{starrypart1, starrypart2} to
model the transits and uses a separate quadratic limb-darkening term for each instrument. Each sector was fit with an independent limb-darkening coefficient. For uninformative limb-darkening priors, we reparameterized the priors following the procedure described in \cite{kipping}. We included a jitter term as a simple noise model for each photometric dataset. We assumed a circular orbit and fix the eccentricity to zero. We also used a dilution term on the photometric model because we want to account for potentially blended background stars in the TESS data. We did not include the dilution term for the RBO data because the higher spatial resolution compared to TESS allows for the star to be isolated from background stars, and our NESSI data confirm that there are no background objects within the RBO point spread function. 

\begin{figure}
    \centering
    \includegraphics[scale=0.27]{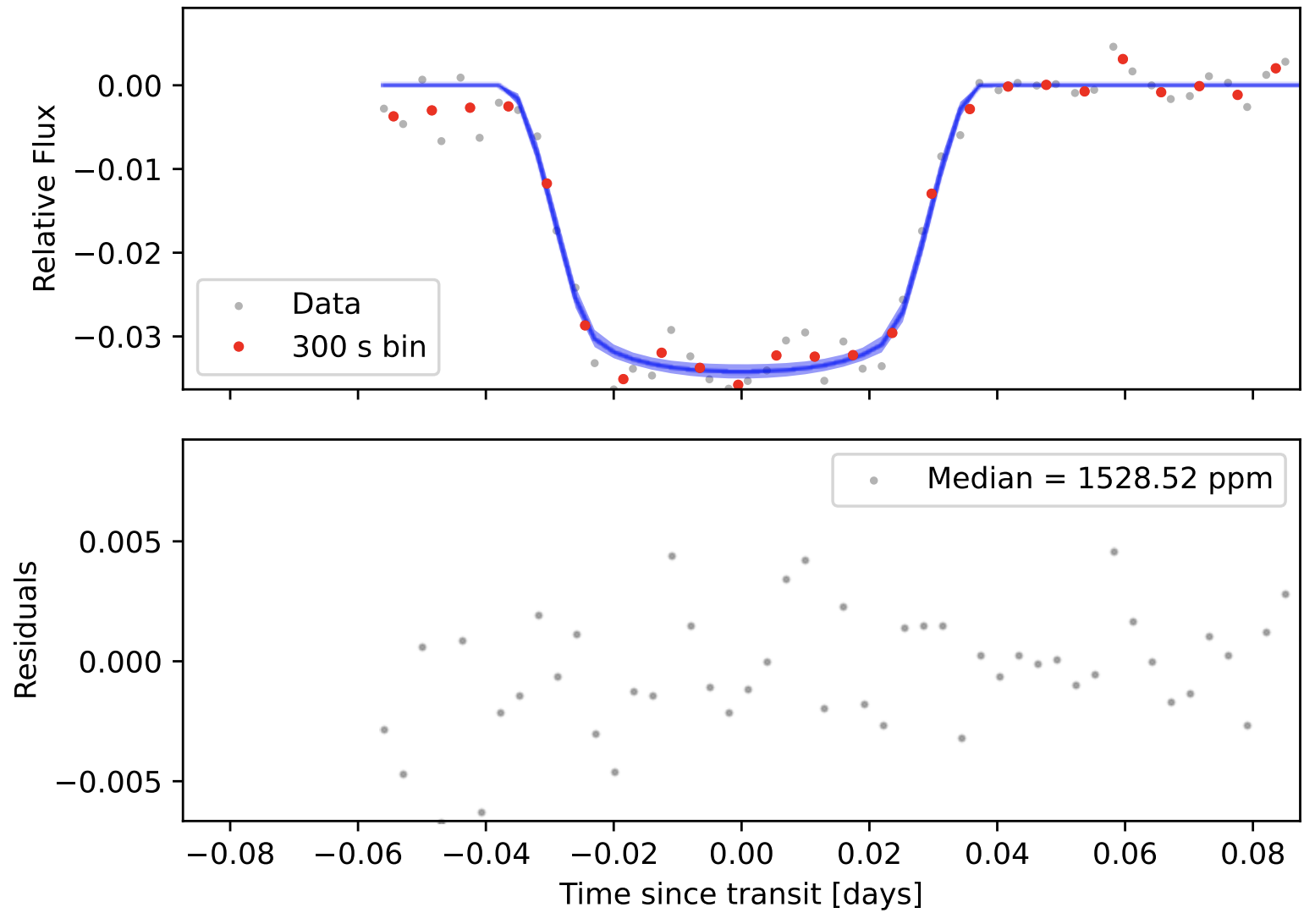}
    \caption{Top: Light curve of RBO data of an exposure time of 240 seconds normalized and detrended using AstroImageJ. MCMC model is overlaid with the blue shaded region indicating a 1$\sigma$ deviation. Bottom: Residuals of this model with a median value of 1528.52 ppm.}
    \label{fig:RBO MCMC}
\end{figure}

We used the standard Keplerian model for the RVs. The photometric model includes the quadratic limb-darkening law \citep{kipping}. We simultaneously fit a Gaussian Process (GP) to the photometric data to detrend the light curve and extracted the transits. Our GP kernel is a mixture of two simple harmonic oscillator terms that can be used to model stellar rotation as a stochastically-driven, damped harmonic oscillator \citep{celerite1, celerite2}. We used this kernel to model the quasi-periodic signal for our likelihood function for the TESS photometry.
We also assumed a linear trend for the RV data.
Figure \ref{fig:RV mcmc total} shows the best fit model overlaid on the RV data with the residuals plotted in the lower panel and shows the phase folded RVs along with the best fit model. We note that the jitter term is relatively high compared to other HPF measurements due to the stellar activity and variability of the star (as seen in Figure \ref{fig:lc}).
Figure \ref{fig:RBO MCMC} shows the RBO photometry with the model overlaid. 
Table \ref{tab:priors} contains a list of our priors used as inputs to $\tt{exoplanet}$.

\subsection{Independent RV Validation}
To test the validity of our joint fit, we used the simplest case of fitting the RV data with $\tt{exoplanet}$.
We adapted the recipe from \cite{exoplanet:single_rv} to fit a single companion using the same RV priors as the joint fit and create a single Keplerian RV model fixing the eccentricity to zero. Our posterior result for the semi-amplitude is 18.26 km/s with a $\sigma$ of 0.17 km/s, which is consistent with our joint fit posterior values.

\subsection{Joint Fit Using Secondary Eclipse Model}

We built upon our initial joint fit model by adding an additional component to model the secondary eclipse in the TESS data. We adopted part of the recipe from \cite{exoplanet_Secondary_eclipsing_recipe}\footnote{\url{ https://gallery.exoplanet.codes/tutorials/eb/}} by including normal priors of the ratios of the of mass ($q$), radius ($R_B/R_A$), and surface brightness ($S$). We applied the secondary eclipse function from $\tt{exoplanet}$ to the TESS sectors to model the secondary eclipse. We did not apply the secondary eclipse function to the RBO data as the duration does not include the secondary eclipse portion of the light curve. The secondary eclipse function models the transits using $\tt{starry}$. As with our initial joint fit, we fixed the eccentricity to zero to improve the stability of the modeling calculation. Solving for eccentricity would be an interesting astrophysical parameter, however, our attempts at allowing this parameter to float caused the model to become unstable. We used two independent quadratic limb-darkening law parameters for the primary star and transiting companion and concluded that our results from this fit are consistent with our single quadratic limb-darkening model. 

\begin{figure*}
    \centering
    \includegraphics[width=\textwidth,height=11cm]{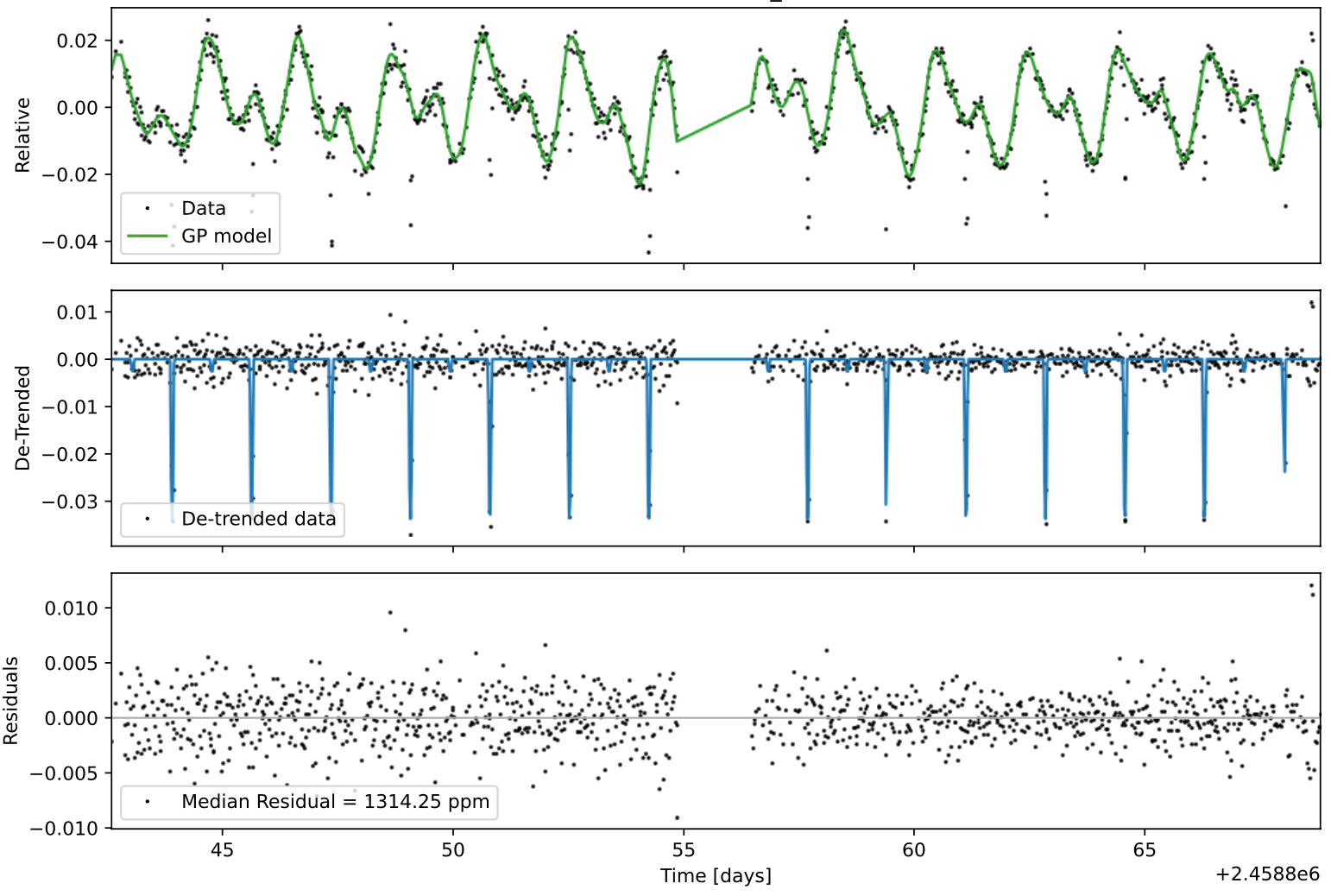}
    \caption{Representative example of TESS Sector 20 photometry along with a stellar rotation GP kernel (top). The detrended photometry is shown in the middle panel, with the eclipses overlaid in blue. The bottom panel shows the residuals.}
    \label{fig:lc model and gp}
\end{figure*}

\begin{figure}[ht]
    \centering
    \includegraphics[scale=0.4]{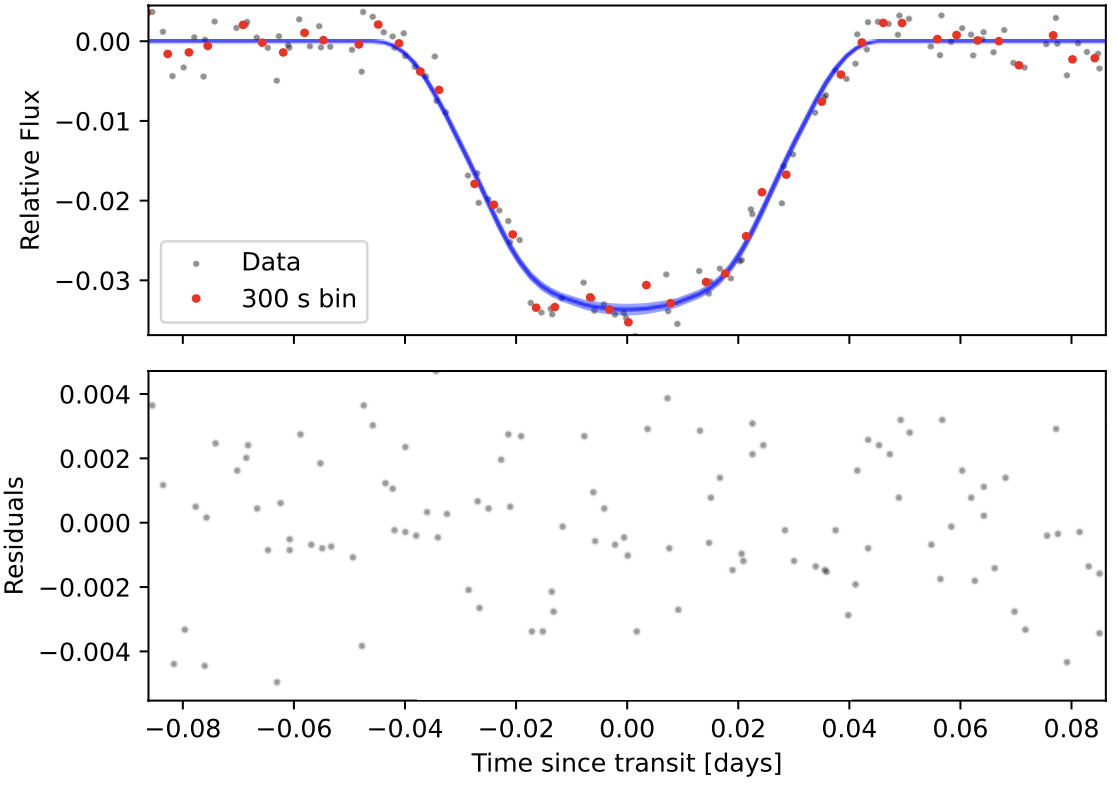}
    \caption{Phase folded photometric observations for TOI-5375 in Sector 20. The grey points are the detrended data, the red points are 300-second bins, and the model is shown in blue with a thin blue shaded region indicating a 1$\sigma$ deviation. The median residual value for the joint fit over all sectors and including the primary and secondary eclipse is 1330 ppm.}
    \label{fig:phase folded model}
\end{figure}

\begin{figure}[ht]
    \centering
    \includegraphics[scale=0.4]{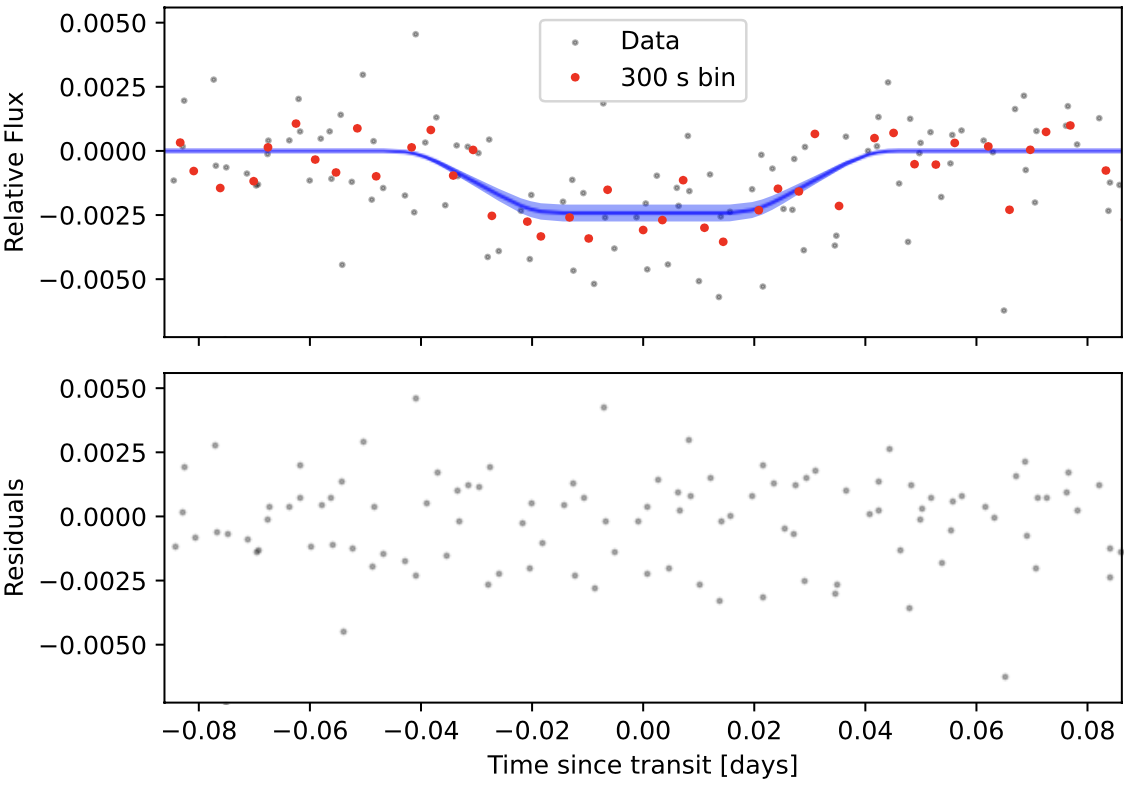}
    \caption{Phase folded photometric observations of the secondary eclipse for TOI-5375 in sector 20. Plot markers are identical to those used in Figure \ref{fig:phase folded model}.}
    \label{fig:LC MCMC secondary}
\end{figure}

Figure \ref{fig:lc model and gp} shows the photometric plot of TESS sector 20 along with a stellar rotation GP kernel. The detrended photometry is shown in the bottom panel along with the optimized mapped eclipses overlaid before running the HMC. The optimized parameter estimates are then used as the initial conditions when running the HMC. Figure \ref{fig:phase folded model} shows the phase folded photometry data from TESS sector 20 with the best fit model posteriors and 1$\sigma$ interval (16th and 84th percentiles).
Figure \ref{fig:LC MCMC secondary} shows the phase folded photometric data of the secondary eclipse from TESS sector 20 with the best fit model and 1$\sigma$ interval. Our analysis yields for the companion a mass of $0.080\pm{0.002} M_{\Sun}$ a radius of 0.1114$^{+0.0048}_{-0.0050} R_{\Sun}$, making the companion, not a planet, but rather a very low mass star, which we designate as TOI-5375 B. Table \ref{tab:posteriors} shows these and other parameters derived from our joint fit analysis.


\begin{deluxetable*}{llc}
\tablecaption{Derived parameters of TOI-5375 using the limb-darkening joint fit model.}
\label{tab:posteriors}
\tablewidth{0pt}
\tablehead{
\colhead{Parameter}&\colhead{Unit}&\colhead{Value}}
\startdata
Parameters:\\
~~Period\dotfill & $P$ (Days)\dotfill & 1.72155391$^{+0.00000144}_{-0.00000142}$\\
~~Impact Parameter\dotfill & $b$ \dotfill& 0.24$^{+0.12}_{-0.14}$\\
~~Semi-amplitude Velocity\dotfill& $K_0$ ($m~s^{-1}$) \dotfill&18256.82$^{+209.84}_{-208.49}$ \\
~~RV trend\dotfill&$dv/dt$ ($m~s^{-1}~yr^{-1}$)\dotfill& 0.14$^{+4.92}_{-5.07}$\\
~~RV jitter\dotfill& $\sigma_{HPF}$ ($m~s^{-1}$)\dotfill & 424.25$^{+134.28}_{-89.55}$\\
~~RV offset\dotfill& $\gamma_{HPF}$ ($m~s^{-1}$)\dotfill &-8535.87$^{+163.89}_{-164.016}$\\
Transit Parameters:\\
~~Transit midpoint\dotfill& $T_0$ ($\mathrm{BJD}_{\mathrm{TDB}}$) \dotfill& $2458843.91098\pm{0.00032}$\\
~~Scaled radius\dotfill& $R_B/R_A$ \dotfill& $0.1493\pm{0.0030}$\\
~~Flux ratio\dotfill&$S$\dotfill&0.086$^{+0.012}_{-0.011}$\\
~~Scaled semi-major axis\dotfill& $a/R_A$ \dotfill& 8.65$^{+0.20}_{-0.25}$\\
~~Eclipse depth\dotfill& $F_{ecl}$ \dotfill& 0.00193$^{+0.00026}_{-0.00025}$\\
~~Inclination\dotfill& $i$ (degrees) \dotfill& 88.41$^{+0.97}_{-0.86}$\\
~~Transit Duration\dotfill & $T_{14}$ (days)\dotfill& 0.07126$^{+0.00097}_{-0.00095}$\\
~~Photometric Jitter\dotfill & $\sigma_{\rm{TESS~S20}}$ \dotfill & 0.001740$^{+0.000081}_{-0.000082}$\\
&$\sigma_{\rm{TESS~S26}}$ \dotfill & 0.001626$^{+0.00013}_{-0.00012}$\\
&$\sigma_{\rm{TESS~S40}}$ \dotfill & $0.001931\pm{0.000093}$\\
&$\sigma_{\rm{RBO}} $ \tablenotemark{a}\dotfill& 0.00195$^{+0.00028}_{-0.00027}$\\
~~Dilution\dotfill& $D_{TESS~S20}$ \dotfill& 0.926$^{+0.033}_{-0.032}$\\
& $D_{TESS~S26}$ \dotfill & 0.821$^{+0.033}_{-0.031}$\\
& $D_{TESS~S40}$ \dotfill & 0.710$^{+0.024}_{-0.023}$\\
~~Rotation Period\dotfill& Prot (days)\dotfill & 1.9716$^{+0.0080}_{-0.0083}$\\
Companion Parameters: \\
~~Radius\dotfill& $R_{B}$ $(R_\Earth)$\dotfill& 10.71$^{+2.057}_{-0.97}$\\
& $R_{B}$ $(R_{J})$\dotfill& 0.96$^{+0.18}_{-0.087}$\\
& $R_{B}$ $(R_\Sun)$\dotfill& 0.098$^{+0.019}_{-0.0089}$\\
~~Mass\dotfill& $M_{B}$ $(M_\Earth)$ \dotfill& $26642.4\pm{666.06}$\\
& $M_{B}$ $(M_{J})$ \dotfill& $84.37\pm{2.02}$\\
& $M_{B}$ $(M_\Sun)$ \dotfill& $0.080\pm{0.002}$\\
~~Temperature\dotfill& $T_B$ (K)\dotfill& $2600\pm{70}$\\
~~Semi-major Axis\dotfill& $a$ (AU) \dotfill& 0.02484$^{+0.00024}_{-0.00026}$\\
\enddata
\tablenotetext{a}{RBO parameters come from joint fit using quadratic limb-darkening function.}
\end{deluxetable*}

\section{Discussion}
\label{sec:Discussion}

Understanding the characteristics of companion objects requires knowledge of the host star. The contrast ratio of exoplanet systems makes secondary eclipse detection nearly impossible. Therefore, deriving the physical parameters for exoplanets is often reliant on the accuracy of stellar evolution models.
Eclipsing binary systems, where the secondary's light can be detected, are important for constraining those stellar evolution models.
In particular, objects near the hydrogen-burning mass limit, like TOI-5375 B, are able to measure the mass and radius mostly independent of models. An estimate of the companion age would indeed make this a benchmark VLMS.

The simplest way to determine the age of the companion is to assume it is coeval with the primary star, for which we can constrain the age. Isochrone models and asteroseismology are less reliable for low-mass stars than for solar-type stars for determining ages, so age estimates for our M-dwarf primary star are weakly constrained at best.
One property of low-mass stars we can exploit is the rotation period. M dwarfs lose angular momentum as they age which results in their rotation period increasing \citep{EngleGuinan2011}. Therefore, rotation periods can be used to estimate the ages of M dwarfs \citep[see;][]{kiraga2007, Guinan2016example, Popinchalk_2021, 2003_barnes}.
\cite{Engle_2018} provide a relation to calculate the age of an early M dwarf (M0-1 V stars) using its rotational period
\begin{equation}
    t(Gyrs) = 0.365+ 0.019*P_{rot}^{1.457} (days).
\label{eq1}
\end{equation}
If we attribute the variability of the host star to evolving star spots over each sector of TESS data, we can use the star spots to extract the rotation period using the period of the GP of 1.9716$^{+0.0080}_{-0.0083}$ days.
We independently measured the rotation period using the periodogram function from lightkurve \citep{2018lightkurve} for each TESS sector and a joint periodogram. The joint periodogram yielded a rotation period of $\sim$1.99 days; this analysis is broadly consistent (2$\sigma$) with the rotation period extracted from the GP fit. This period is also visually consistent with the 4\% modulation seen in Figure \ref{fig:lc}.
Equation \ref{eq1} suggests a rotation period derived age of $\sim$400 Myrs. This value is consistent within 1$\sigma$ of the expected age of an early M dwarf with a rotation period between 1$<$P$<$10 days as seen in \cite{Newton_2016}. However, we note that due to the complex nature of the close binary system, which has potentially significant tidal effects affecting the angular momentum of the system, rotation-based ages derived for single stars and widely separated binaries may not apply.

The depth of the secondary eclipse observed in TESS can be modeled as a function of various fundamental properties \citep[e.g.,][]{Charbonneau, Esteves2013, Shporer2017},
\begin{equation}
    Depth = \left(\frac{R_2}{R_\ast}\right)^2 \frac{\int \tau(\lambda)F_{2,\nu}(\lambda, T_2)d\lambda}{\int \tau(\lambda)F_{\ast,\nu}(\lambda, T_e)d\lambda} + A_g\left(\frac{R_2}{a}\right)^2,
\end{equation}
where $\tau(\lambda)$ is the TESS transmission function, $T_e$ and 
$F_{\ast,\nu}$($\lambda$, $T_e$) are the effective temperature and flux of the host star, $T_2$ and $F_{2,\nu}$($\lambda$, $T_2$) is the brightness temperature and flux of TOI-5375 B, and $A_g$ is the geometric albedo. For TOI-5375 B, we ignored any contribution to the eclipse depth from reflected light and ellipsoidal variations \citep[e.g.,][]{Shporer2017}. 
We used our posterior distribution to estimate the fluxes of the host star and companion using BT-Settl models \citep{Allard} based on the \cite{2011Caffau} solar abundances. We used $\tt{SPISEA}$ \citep{SPISEA} , an open-source python package that simulates simple stellar populations, as an interface to the BT-Settl model grid. This method yields a temperature of $2600\pm{70}$ K, which is consistent with TOI-5375 B being a late-M-type VLMS.

We used the Bayesian Analysis for Nearby Young AssociatioNs $\Sigma$ (BANYAN$\Sigma$) to calculate the membership probability of TOI-5375 with any nearby stellar clusters within 150 pc of the Sun \citep{Gagn2018}. BANYAN$\Sigma$ uses multivariate Gaussian models in 6-dimensional space on 27 young associations with ages in the range $\sim$1–800 Myr. Using the coordinates, proper motion, radial velocity, and parallax from the GAIA DR3 archive \citep{GAIA_2016, GAIA_2022}, BANYAN$\Sigma$ indicates a 99.9\% likelihood of TOI-5375 being associated with the field. 

\begin{figure*}
    \centering
    \includegraphics[width=\textwidth,height=12cm]{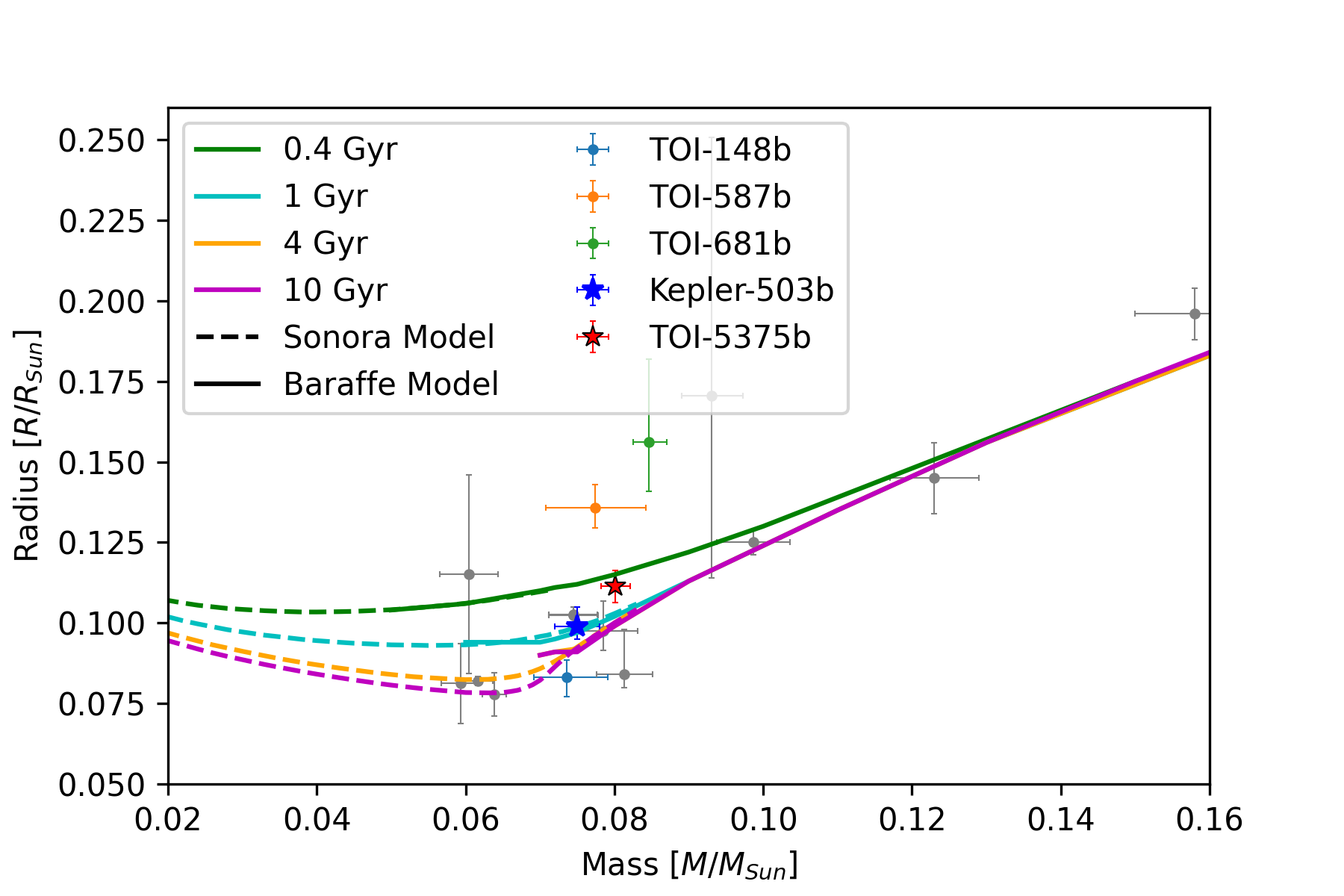}
    \caption{Derived mass and radius of the companion of TOI-5375 plotted with other objects near the hydrogen-burning mass limit. For reference, Kepler-503 is plotted in blue \citep{Caleb}; TOI-148, TOI-587, and TOI-681 are plotted in blue, orange, and green, respectively, while other low-mass stars and high-mass substellar companions from \cite{BD_references} as gray circles. We show solar metallicity ([M/H] = 0.0) evolutionary isochrone tracks from \cite{Baraffe} (solid lines) and from \citep{Sonora} (dashed lines). These models are for substellar companions and low-mass stars, and span the ages 0.4 1, 4, and 10 Gyr.}
    \label{fig:isochrone}
\end{figure*}

Figure \ref{fig:isochrone} shows TOI-5375 B plotted on a mass-radius distribution of substellar and other low-mass stars near the hydrogen-burning mass limit. We also show the solar metallicity ([M/H] = 0.0) evolutionary isochrone tracks from Baraffe \citep{Baraffe} and Sonora \citep{Sonora}. Our mass-radius results are consistent with the 0.4 Gyr model, and this is consistent with our rotation-based estimate of the age of the system.
However, in this region of parameter space, isochrones corresponding to older ages begin to fall on top of each other as the stars settle on the main sequence, so at 2$\sigma$ our radius measurement is consistent with a broad range of isochrone ages.
TOI-5375 B is comparable in mass and radius to Kepler-503b \citep{Caleb} although Kepler-503b's age is much older at $\sim$6.7Gyrs. It is gratifying to see that both objects are consistent with the isochrone tracks for their respective ages.

\subsection{Additional Observations}
Our $\tt{HPF-SpecMatch}$ analysis measured a spectroscopic $v\sin i_A$ of $16.7\pm{0.9}$km/s. 
Combining this with the 1.9716$^{+0.0080}_{-0.0083}$ day rotation period from our joint fit, and stellar radius of $0.632\pm{0.019}R_{\Sun}$ allows an estimate of the stellar inclination. Using the methodology from \cite{2020_masuda_winn}, and allowing inclination to range from 0 to 180 degrees, yields a stellar inclination estimate of $90\pm{13}$ degrees. Our modeled orbital inclination posterior is 88.41$^{+0.97}_{-0.86}$ degrees. Together, our joint fit, the $v\sin i_A$, and rotation period suggest both the stellar equator and orbit of TOI-5375 B are close to edge on, and most likely well aligned. 
Independent measurements of the obliquity using the Rossiter-McLaughlin (RM) effect \citep{Triaud2018} would directly confirm these results. However, due to the relative faintness and length of transit duration (1.74 hours), acquiring RV data from the HET would be nearly impossible, so a different spectrograph, such as MAROON-X \citep{MaroonX} at Gemini would be required.

The modulation seen in the different TESS sectors in Figure \ref{fig:lc} could be used to deduce the atmospheric circulation \citep[e.g.,][]{atmosphere_example} and future efforts could explore the efficacy of heat circulation in the companion based on the temperature measured in transit and in the eclipse position, however such an analysis is beyond the scope of this paper.

\section{Summary}
\label{sec:summary}
We present ground-based follow up data from HPF, NESSI, and RBO, and use it along with TESS photometry to carry out a Hamiltonian Markov Chain (HMC) joint fit that characterizes the companion to TOI-5375. This analysis shows TOI-5375 B is a VLMS with a mass of
$0.080\pm{0.002} M_{\Sun}$ a radius of 0.1114$^{+0.0048}_{-0.0050} R_{\Sun}$, and brightness temperature of $2600\pm{70}$ K, on a 1.721553$^{+0.000001}_{-0.000001}$ day orbit. The host star has a rotation period of 1.9716$^{+0.0080}_{-0.0083}$ days, determined by the spot-induced periodicity in the lightcurve. The rotation period is suggestive of an age of $\sim$400 Myrs measured using single star evolution of wide binaries and is not associated with any nearby clusters. TOI-5375 is amenable to additional modeling including atmospheric circulation, RM observations to measure obliquity, and the 3-D architecture of the orbit.

\section{Acknowledgments}*
We acknowledge support from NSF grants AST-1006676, AST-1126413, AST-1310885, AST-1310875, ATI 2009554, ATI 2009889, ATI-2009982, AST-2108512, and the NASA Astrobiology Institute (NNA09DA76A) in the pursuit of precision radial velocities in the NIR. The HPF team also acknowledges support from the Heising-Simons Foundation via grant 2017-0494. CIC acknowledges support by NASA Headquarters through an appointment to the NASA Postdoctoral Program at the Goddard Space Flight Center, administered by USRA through a contract with NASA and the NASA Earth and Space Science Fellowship Program through grant 80NSSC18K1114. GS acknowledges support provided by NASA through the NASA Hubble Fellowship grant HST-HF2-51519.001-A awarded by the Space Telescope Science Institute, which is operated by the Association of Universities for Research in Astronomy, Inc., for NASA, under contract NAS5-26555.

Funding for the TESS mission is provided by NASA’s Science Mission directorate. 
The Hobby-Eberly Telescope is a joint project of the University of Texas at Austin, the Pennsylvania State University, Ludwig-Maximilians-Universität München, and Georg-August Universität Gottingen. The HET is named in honor of its principal benefactors, William P. Hobby and Robert E. Eberly. The HET collaboration acknowledges the support and resources from the Texas Advanced Computing Center. We thank the Resident Astronomers and Telescope Operators at the HET for the skillful execution of our observations with HPF. We would like to acknowledge that the HET is built on Indigenous land. Moreover, we would like to acknowledge and pay our respects to the Carrizo \& Comecrudo, Coahuiltecan, Caddo, Tonkawa, Comanche, Lipan Apache, Alabama-Coushatta, Kickapoo, Tigua Pueblo, and all the American Indian and Indigenous Peoples and communities who have been or have become a part of these lands and territories in Texas, here on Turtle Island.

Data presented herein were obtained at the WIYN Observatory from telescope time allocated to NN-EXPLORE through the scientific partnership of the National Aeronautics and Space Administration, the National Science Foundation, and the National Optical Astronomy Observatory. WIYN is a joint facility of the University of Wisconsin–Madison, Indiana University, NSF’s NOIRLab, the Pennsylvania State University, Purdue University, University of California, Irvine, and the University of Missouri. NESSI was funded by the NASA Exoplanet Exploration Program and the NASA Ames Research Center. NESSI was built at the Ames Research Center by Steve B. Howell, Nic Scott, Elliott P. Horch, and Emmett Quigley. NSF’s NOIRLab, managed by the Association of Universities for Research in Astronomy (AURA) under a cooperative agreement with the National Science Foundation. The authors are honored to be permitted to conduct astronomical research on Iolkam Du’ag (Kitt Peak), a mountain with particular significance to the Tohono O’odham. Deepest gratitude to Zade Arnold, Joe Davis, Michelle Edwards, John Ehret, Tina Juan, Brian Pisarek, Aaron Rowe, Fred Wortman, the Eastern Area Incident Management Team, and all of the firefighters and air support crew who fought the recent Contreras fire.

An allocation of computer time from the UA Research Computing High Performance Computing (HPC) at the University of Arizona and the prompt assistance of the associated computer support group is gratefully acknowledged.

The Center for Exoplanets and Habitable Worlds is supported by Penn State and the Eberly College of Science.
The Pennsylvania State University campuses are located on the original homelands of the Erie, Haudenosaunee (Seneca, Cayuga, Onondaga, Oneida, Mohawk, and Tuscarora), Lenape (Delaware Nation, Delaware Tribe, Stockbridge-Munsee), Shawnee (Absentee, Eastern, and Oklahoma), Susquehannock, and Wahzhazhe (Osage) Nations. As a land grant institution, we acknowledge and honor the traditional caretakers of these lands and strive to understand and model their responsible stewardship. We also acknowledge the longer history of these lands and our place in that history.

Some of the data presented was obtained from MAST at STScI. Support for MAST for non-HST data is provided by the NASA Office of Space Science via grant NNX09AF08G and by other grants and contracts. This work includes data collected by the TESS mission, which are publicly available from MAST. Funding for the TESS mission is provided by the NASA Science Mission directorate.

This work has made use of data from the European Space Agency (ESA) mission
{\it Gaia} (\url{https://www.cosmos.esa.int/gaia}), processed by the {\it Gaia}
Data Processing and Analysis Consortium (DPAC,
\url{https://www.cosmos.esa.int/web/gaia/dpac/consortium}). Funding for the DPAC
has been provided by national institutions, in particular the institutions
participating in the {\it Gaia} Multilateral Agreement.

We thank the anonymous referee for insightful comments that have improved the quality of this work.

\facilities{TESS, RBO, WIYN (NESSI), HET (HPF)}

\software{
Astropy              \citep{astropy1, astropy2},
lightkurve           \citep{2018lightkurve},
Matplotlib           \citep{matplotlib},
NumPy                \citep{numpy},
pandas               \citep{pandas},
SciPy                \citep{scipy1, scipy2}
}

\bibliography{references.bib}{}
\bibliographystyle{aasjournal}

\end{document}